\documentclass[a4paper,10pt,journal,twoside]{IEEEtran}
\usepackage{cite}
\usepackage{amsmath,amssymb,amsfonts}
\usepackage{algorithmic}
\usepackage{graphicx}
\usepackage{textcomp}
\usepackage{paralist}
\usepackage{enumitem}        % Listas personalizables
\usepackage{flushend}        % Para equilibrar las columnas al final

\usepackage{float}
\usepackage{caption}
\usepackage{subcaption}

\usepackage{comment} % para comentar trozos grandes sin borrarlos

\usepackage[toc,nopostdot, nonumberlist,style=long,automake,acronym]{glossaries}

\usepackage[dvipsnames]{xcolor}
\usepackage{lipsum}

\usepackage{siunitx}
\usepackage[normalem]{ulem} % Para poder tachar cosas cuando corregimos
\usepackage{acronym} 
\DeclareSIUnit \voltampere { VA } %apparent power 
\DeclareSIUnit \var { var } %volt-ampere reactive - idle power
\DeclareSIUnit \pu {pu} % per-unit

\makeglossaries % Comment if no glossaries are needed
\newacronym{CIG}{CIG}{converter-interfaced generation}
\newacronym{DFIG}{DFIG}{doubly-fed induction generator}
\newacronym{DOF}{DOF}{degree-of-freedom}
\newacronym{GFL}{GFL}{grid-following}
\newacronym{GFM}{GFM}{grid-forming}
\newacronym{GSC}{GSC}{grid-side converter}
\newacronym{PCC}{PCC}{point of common coupling}
\newacronym{PI}{PI}{Proportional-Integral}
\newacronym{PLL}{PLL}{phase-locked loop}
\newacronym{pu}{pu}{per unit}
\newacronym{RES}{RES}{renewable energy sources}
\newacronym{RoCoF}{RoCoF}{rate of change of frequency}
\newacronym{RoCoV}{RoCoV}{rate of change of voltage}
\newacronym{RSC}{RSC}{rotor-side converter}
\newacronym{SCR}{SCR}{short-circuit ratio}
\newacronym{VSC}{VSC}{voltage-sorce converter}
\newacronym{VSM}{VSM}{virtual synchronous machine}

\def\BibTeX{{\rm B\kern-.05em{\sc i\kern-.025em b}\kern-.08em
    T\kern-.1667em\lower.7ex\hbox{E}\kern-.125emX}}
\newcommand{\Alpha}{\mathrm{A}}
\newcommand{\Beta}{\mathrm{B}}

\begin{document}

%Frequency-Domain 
\title{Multi-Loop Design of Virtual Synchronous Machine Control for DFIG-Based Wind Farms}
%\author{Javier Garc\'ia-Aguilar, \IEEEmembership{Student, IEEE}, Aurelio Garc\'ia-Cerrada, \IEEEmembership{SM, IEEE}, Juan Luis Zamora,  Emilio Bueno and Elena Saiz 
\author{Javier Garc\'ia-Aguilar, Aurelio Garc\'ia-Cerrada, Juan L. Zamora,  Emilio Bueno,  Elena Saiz, Almudena Mu\~{n}oz-Babiano and Mohammad E. Zarei
\thanks{This work was partially supported by the "\textsc{Asociación/Colegio Nacional de Ingenieros del ICAI}", the European Regional Development Fund, The Spanish Ministry of Science, Innovation and Universities and the Spanish State Research Agency with reference RTC-2017-6074-3 in collaboration with Siemens Gamesa, and Siemens Gamesa Renewable Energy Innovation \& Technology S.L. {\em IIT Working Paper, ref. IIT-25-314WP}}
\thanks{J. Garc\'{\i}a-Aguilar, A. Garc\'{\i}a-Cerrada and J.L. Zamora are with Institute for Research in Technology, Comillas Pontifical University, Madrid, Spain. (emails: [jgaguilar,aurelio,zamora]@comillas.edu)}
\thanks{E. Bueno is with Alcal\'{a} University, Alcal\'{a} de Henares, Madrid. (email: emilio.bueno@uah.es)}
\thanks{E. Saiz, A. Mu\~{n}oz-Babiano and M.E. Zarei are with Siemens Gamesa Renewable Energy Innovation \& Technology S.L, Madrid. (email:[elena.saiz,almudena.munoz,mohammad.zarei]@siemens-energy.com)}
\thanks{{\bf Note:} An updated version of this manuscript will be published in the Journal of Modern Power Systems and Clean Energy with doi: {10.35833/MPCE.2025.000640}}}

\maketitle

% Abstract and keywords
\vspace{-5mm}
\begin{abstract}
The displacement of synchronous generators by converter–interfaced renewable energy sources obliges wind farms to provide inertia, damping, and voltage support, above all in increasingly weak grid conditions.  
This paper presents a \emph{co-ordinated frequency-domain methodology} for tuning all control layers of doubly-fed induction generators (DFIGs) within a wind farm operated as a Virtual Synchronous Machine (VSM).  
Starting from a full small-signal linearisation that preserves loop-to-loop and machine-to-machine couplings, the procedure reshapes every local open loop to explicit phase-margin targets through a single, prioritised iteration. The resulting controllers provide a step response and stability margins close to those programmed at the design stage, in spite of the cross coupling between control loops. Since controller synthesis relies exclusively on classical loop-shaping tools available in commercial simulation suites, it is readily applicable to industrial-scale projects.  
\end{abstract}


\glsresetall

\vspace{-3mm}
\section{Introduction}
\label{sec:intro}

Type-III wind generators based on \glspl{DFIG} currently dominate installed wind capacity (mainly onshore) because they enable variable-speed operation with reduced cost~\cite{Abad2011}. %their partially-rated back-to-back rotor converter reduces cost while enabling variable-speed operation~\cite{Abad2011}.
Traditionally, \gls{GFL}  technology has dominated \gls{DFIG}-based generation. It uses a \gls{PLL} to synchronise with the grid; an inner rotor-side current loop controls electromagnetic torque, while outer loops on the grid-side converter regulate DC-link voltage and stator-terminal reactive power~\cite{Jacob2015}.
%\cite{Morren2006,Jacob2015}.
Under strong-grid conditions, this arrangement works well because loop interactions are weak, and the voltage at the \gls{PCC} is not affected much by the actions of the \gls{DFIG}.
However, in weak grids or parallel \gls{DFIG} operation, the arrangement described is prone to adverse cross-couplings and \gls{PCC} voltage variations that can excite torsional and sub-synchronous modes~\cite{Wen2018}.

\Gls{GFM} control has emerged as a key enabler for a high penetration of \gls{CIG} in general~\cite{Rocabert2012} and is now being requested for \gls{DFIG}-based generators. 
Among \gls{GFM} strategies, the \gls{VSM} emulates the swing equation and electromagnetic behaviour of a synchronous generator, naturally providing the so-called virtual inertia and damping~\cite{Zhong2011}. Recent publications 
have shown that \glspl{DFIG} can also be operated in \gls{GFM} mode without jeopardising their cost advantages~\cite{GonzalezCajigas2023}.
Most published implementations, however, simply \emph{overlay} a \gls{VSM} power loop on the inner current and voltage loops, assuming the inner layers work seamlessly. Unfortunately, this hypothesis breaks down for low \gls{SCR} ($\mathrm{SCR}\le 2$) or when multiple \glspl{DFIG} share a collector feeder~\cite{Navarro2019,TaskForce2023}.  In these cases, the impact of the time response of the inner loops on the outer loops cannot be underestimated. 
Small-signal and impedance-based analyses have revealed the root causes of \gls{GFM}–\gls{DFIG} instabilities.
The aggressive power-angle modulation by the \gls{VSM} outer loop is affected by the limited (although fast) speed of the rotor current controllers, especially when the \gls{PLL} or stator-flux observer introduces additional delay~\cite{Sun2011,Li2022,Yang2023}.
The problem is exacerbated by unequal collector impedances in multi-machine systems and power line resonances~\cite{Culibrk2022}.

In this regard, recent literature looks at co-ordinated tuning of all relevant loops —current, DC-link, power/voltage and \gls{PLL}— in all forms of \gls{CIG} (including \gls{DFIG}-based generation systems). In fact, important organisations with responsibilities in energy systems such as IEEE~\cite{IEEE-Interoperability2022}, ENTSO-E~\cite{24-ENTSOE-1-2022}, and CIGRE~\cite{CIGRE-Multi} have warned of dangerous interactions between converter controllers and between classical synchronous generators and converter controllers. The detection and analysis of these interactions, and countermeasures to prevent system malfunction, are of special concern to these organisations.

Sophisticated techniques like $H_\infty$ optimisation and $\mu$-synthesis offer robust control design frameworks and have been applied to single converters~\cite{Chen2020,Wang2016} and extended to multi-converter systems~\cite{Chaudhuri2014,VanCutsem2015,HInfinityMu_GFM_Optimization2024}. However, they often result in high-order controllers and can be computationally intensive for large multi-machine systems, potentially obscuring a direct physical insight for tuning. Similarly, time-domain optimisation methods, while capable of handling non-linearities and complex constraints~\cite{TimeDomainRobustControlGFM2024}, can be sensitive to model accuracy and may struggle with guaranteeing specific frequency-domain performance metrics like phase and gain margins across interacting loops.
Multivariable techniques from the process-control community have also been adapted to wind turbines~\cite{Skogestad2005,Johansson2014} but without continuation.

This paper proposed a transparent and scalable methodology based on  classical frequency-domain loop-shaping techniques for the design of \gls{VSM}-controlled \gls{DFIG} wind farms. Unlike recent automated $H_\infty/\mu$ tuning, the proposed methodology gives direct control over individual loop characteristics (gain/phase margins) through an iterative, physically-grounded process. This should ease interpretation and adjustment by practising engineers. %unlike other methods that, while mathematically optimal, might act more like ``black boxes.''
\begin{comment}
% INTERESANTE PERO REPITE LAS COSAS
In fact, a practical methodology that
\begin{enumerate}[label=(\alph*)]
  \item retains all loop-to-loop and machine-to-machine couplings in a tractable model;\vspace{2pt}
  \item supplies clear gain- and phase-margin targets; and\vspace{2pt}
  \item scales to multi-machine wind farms without prohibitive computational effort, offering a clear advantage in terms of implementation complexity and intuitive tuning over some advanced robust control synthesis techniques,
\end{enumerate}
has not been presented yet.
\end{comment}

The main contributions of this paper can be summarised as follows: 
\begin{itemize}
  \item A methodology based on small-signal linearisation that captures the interactions among \gls{DFIG} control loops.
  \item A prioritised single-iteration loop-shaping procedure that prevents the deterioration of  open-loop frequency response stability margins due to controller interactions, even for low SCR 
   and unequal collector impedances for individual \glspl{DFIG}.
  \item The proposed methodology relies solely on classical loop-shaping tools available in commercial simulation suites and is thus directly transferable to industry practice.
\end{itemize}

The main contributions were validated by detailed simulation on a four-machine, 8.4~MW benchmark. 

The remainder of the paper is organised as follows.
Section~\ref{sec:VSM_DFIG_WF} describes the \gls{VSM}-controlled \gls{DFIG} wind-farm benchmark and the modelling assumptions.
Section~\ref{sec:initialControlAnalysis} summarises the initial sequential tuning together with its small-signal assessment.  Section~\ref{sec:interactions} illustrates the loop-to-loop couplings that motivate a coordinated redesign.
Section~\ref{sec:designExample} presents the proposed frequency-domain, multi-loop synthesis procedure and illustrates its application step-by-step.
Section~\ref{sec:results} validates the redesigned controllers through both small-signal and large-signal tests, and Section~\ref{sec:conclusions} concludes the study and outlines directions for future work. 
%=========================================================
%  II. VSM-CONTROLLED DFIG WIND FARM
%=========================================================
  \vspace{-5mm}
  \section{VSM-Controlled DFIG Wind Farm}
\label{sec:VSM_DFIG_WF}

\begin{figure}[!t] 
    \centering
    \includegraphics[height=5cm,width=\linewidth]{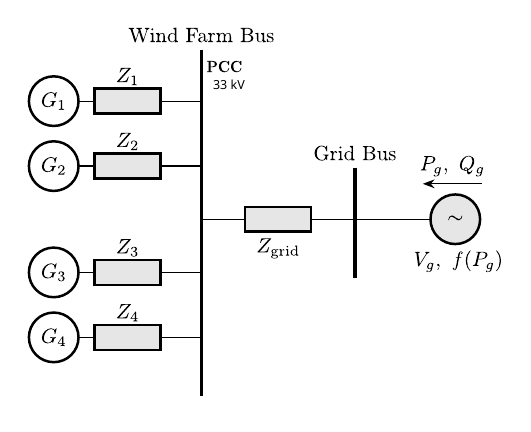}
    \caption{Single-line diagram of the \gls*{VSM}-controlled DFIG wind-farm system.}
    \label{fig:sysOneLine}
    \vspace{-5mm}
\end{figure}

Figure~\ref{fig:sysOneLine} shows the benchmark system analysed in this work. It consists of four identical Type-III wind turbines, based on \glspl{DFIG}, rated at \SI{2.1}{\mega\watt} each, and connected in parallel to the transmission grid via \SI{33}{\kilo\volt} collector feeders and a step-up transformer.%\footnote{The absolute rating is irrelevant for small-signal analysis as all magnitudes are normalised to the bases listed in Appendix~\ref{apx:system-bases}.} % donde están las bases, habrá que decirlo luego

The collector impedances are intentionally non-uniform: \(Z_{1}=Z_{2}=0.06\)~pu, whereas  \(Z_{3}=Z_{4}=0.12\)~pu. These per-unit values aggregate the respective cable and switchgear contributions of each feeder line, and the transformer of each DFIG. The \gls{PCC} itself is linked to an equivalent grid node via a lumped Thevenin impedance ($Z_{grid}$) so that, at the \gls{PCC}, \(\mathrm{SCR}=1\) (this SCR is calculated using a base power equal to nominal of the plant 
with four DFIGs). This models an extremely weak grid, %a condition known to exacerbate controller interactions and stability issues. 
with a colourful arrangement, while maintaining a tractable overall model for detailed frequency-domain analysis.

Unless otherwise stated, all equations, and electrical and mechanical variables are expressed in \gls{pu} on the bases detailed in Appendix~\ref{apx:system-bases}. Network and passive elements are represented in a synchronous $dq$ reference frame (also as complex numbers). The $d$-axis of this frame (which establishes the global $dq$ reference frame for the simulation of the whole system) is aligned with the grid-voltage space vector using the power-invariant Park transformation. Therefore $\vec{v}_g = 1 \angle{0^{\rm o}}$.  Subscripts $(\cdot)_{d}$ and $(\cdot)_{q}$ denote $d$-axis and $q$-axis components, respectively, for all vector quantities. 

%---------------------------------------------------------
  \vspace{-5mm}
\subsection{Grid model}\label{subsec:grid_model}
\begin{figure}[t] % Figura original, sin cambios
  \centering
  \includegraphics[width=\linewidth]{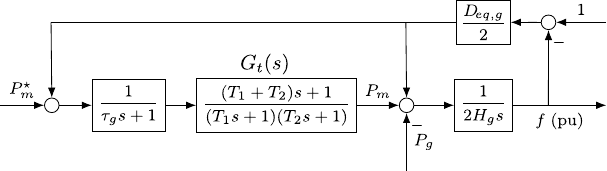}
  \caption{Simplified grid-frequency dynamics with aggregate damping \(D_{\!eq,g}\) and inertia \(H_g\).}
  \label{fig:grid-frec-b}
  \vspace{-5mm}
\end{figure}

%---------------------------------------------------------
The external AC system is modelled with fixed RMS voltage magnitude $\lvert v_{g}\rvert = 1$~pu, but with a time-varying frequency $f$ (in pu):
\begin{equation}
  f\;(\text{pu}) = \omega_{g}/\omega_b,
  \qquad
  \Delta f \triangleq f-1 , \label{eq.freqinpuAGC}
\end{equation}
where \(\omega_{g}\) (rad/s) is the instantaneous grid angular frequency, and \(\omega_b = 2\pi\!\times\!50\)~rad/s is its nominal (and base) value. Figure~\ref{fig:grid-frec-b} depicts the simplified dynamic model of the external grid, where the feedback gain \(D_{\!eq,g}/2\) serves a dual role:
\begin{itemize}
    \item[(a)] In the mechanical branch (representing governor-turbine response), the term \(-(D_{\!eq,g}/2)\,\Delta f\) provides the \emph{primary frequency-control action}. Equating this to the classical static droop characteristic \(\Delta P = -\Delta f/R_{\text{eq}}\), the \emph{equivalent droop} of the external system is \(R_{\text{eq}} = 2/D_{\!eq,g}\).
    \item[(b)] In the electrical branch, the same feedback term summarises the  damping due to the natural frequency dependence of the connected power demand.
\end{itemize}

Consequently, the simplified time-domain equations governing the grid frequency dynamics are:
\begin{align}
  \tau_g \,\frac{dP_m}{dt} &= -P_m + P_m^{\star}
      - \frac{D_{\!eq,g}}{2}\,\Delta f ,         \label{eq:gov_simpl}\\[4pt]
  2H_g \,\frac{d\Delta f}{dt} &= P_m - P_g
      - \frac{D_{\!eq,g}}{2}\,\Delta f ,         \label{eq:swing_simpl}
\end{align}
where: \(P_m^{\star}\) (pu) is the mechanical power set point of the equivalent generator;
  \(P_m\) (pu) is the mechanical power delivered after the governor–turbine train dynamics; \(P_g\) (pu) is the electrical power exported from the wind farm to the grid at the \gls{PCC}; \(\tau_g\) (s) is the governor time constant of the equivalent generator; and \(H_g\) (s) is the lumped inertia constant of the external system. The second-order turbine dynamics of the grid $G_{t}(s)$ is shown in Fig.~\ref{fig:grid-frec-b} and has been incorporated into the overall state-space model, although omitted from equation (\ref{eq:gov_simpl}) for conciseness ($P_m^* = P_m$).

If \(\omega_i\) (rad/s) is the instantaneous electrical angular speed of any device \(i\) (such as a generator stator, its rotor reference frame, or a grid-side converter internal frame) and \(\delta_i\) (rad) is its electrical angle measured with respect to this common \(dq\) frame, the fundamental kinematic relation between the two frames is then:

  \vspace{-4mm}
\begin{equation}
  \delta_i(t) = \delta_i(t_0) + \!\!\int_{t_0}^{t}
                  \overbrace{\bigl[\omega_i(\tau)-\omega_{g}(\tau)\bigr]}^{d\delta_i(t)/dt}\,\mathrm d\tau,
                  \label{eq:angle_int}
\end{equation}
where \(\omega_{g}(t)\) (rad/s) is the instantaneous angular frequency of  the common reference frame, and \(t_0\) is the initial simulation time. For the nominal operating point, \(\omega_{g}(t_0)=\omega_b = 2\pi\times50\)~rad/s, and by definition, \(\delta_g(t) \equiv 0\). %Recall that the \gls{pu} frequency defined in (\ref{eq.freqinpuAGC}) is \(f = \omega_{g}/\omega_b\).

%---------------------------------------------------------
\vspace{-2mm}
\subsection{Generator model}\label{subsec:gen_model}
%---------------------------------------------------------
The 2-pole \gls{DFIG} model in $pu$ of each of the four generation units follows~\cite{Abad2011}:
%\cite{Krause2013}: 
\begin{eqnarray}
 \vec{v}_{dq s} & = & r_s  \vec{i}_{dq s} + j\omega_s \vec{\psi}_{dqs} + \frac{1}{\omega_b} \frac{d \vec{\psi}_{dq s}}{dt} \label{eq.vstatoralphabetabis} \\
  \vec{v}_{dq r} & = & r_r    \vec{i}_{dq r} +   j(\omega_s-\omega_m) \vec{\psi}_{dqr} +  \frac{1}{\omega_b} \frac{d \vec{\psi}_{dq  r}}{dt} \label{eq.vrotoralphabetabis}
\end{eqnarray}
where~(\ref{eq.vstatoralphabetabis}) contains stator-related space vectors: voltage ($\vec{v}_{dqs}$), current ($\vec{i}_{dqs}$) and flux linkage ($\vec{\psi}_{dqs}$) and~(\ref{eq.vrotoralphabetabis}) contains rotor-related space vectors; $\omega_s$ is the angular speed of the reference frame synchronous with the stator variables and $\omega_m$ is the rotor angular speed. In addition: 
\begin{equation}
    \vec{\psi}_{dqs} = L_{s} \vec{i}_{dqs} + L_M \vec{i}_{dqr} \;\;\& \;\; \vec{\psi}_{dqr} = L_{M} \vec{i}_{dqs} + L_r \vec{i}_{dqr} \label{eq.flujod}  %\\
%    \vec{\psi}_{dqr} &=& L_{M} \vec{i}_{dqs} + L_r \vec{i}_{dqr} \label{eq.flujoq}
\end{equation}
and the electric torque applied to the shaft is ($T_e$ applied in the positive direction of $\omega_r$): 
\begin{equation}
T_e = \frac{1}{\sqrt{3}} L_M {\rm Imag} \left(\vec{i}_{dqs} \vec{i}_{dqr}^*\right) \label{eq.torque}
\end{equation}

The wind-turbine blades were assume to work with a pitch angle $\beta =1^{{\rm o}}$ and the turbine power coefficient was approximated  by: 
\begin{eqnarray}
C_p(\lambda) & = & 
-0.4958
+0.2776\,\lambda
-0.02561\,\lambda^{2}
+ \nonumber \\ 
\null & + & 8.7047\times10^{-4}\,\lambda^{3}
-1.1331\times10^{-5}\,\lambda^{4}
\label{eq.Cpinnumbersagc}
\end{eqnarray}
with $\lambda$ being the usual ratio between the blade-tip speed and the wind speed. With this characteristics, $C_p$ reaches of maximum value of nearly $0.5$ when $\lambda \approx 9$. 

The mechanical power extracted from the wind can then be calculated as ($\rho$ is the air density, $R$ is the blade radius and $v$ is the wind speed): 
\begin{equation}
    P_{mec} = \frac{1}{2} \rho \pi R^2 C_p(\lambda) v^3 \label{eq.Pmecagc}
\end{equation}

The electromechanical drivetrain for each turbine-generator set was simulated as the classical two-mass model: 

\begin{eqnarray}
    \frac{d\omega_t}{dt} &= &\frac{1}{2H_t}\left[T_m-D_t\omega_t - T_{tg} - D_{tg}(\omega_t - \omega_r)\right] \label{eq.omega1masas}\\
    \frac{dT_{tg}}{dt} &=&\omega_o K_{tg}(\omega_t - \omega_r) \label{eq.parmasas}\\
    \frac{d\omega_r}{dt} &= &\frac{1}{2H_g}\left[T_{tg}+T_e-D_q\omega_r + D_{tg}(\omega_t-\omega_r)\right] \label{eq.omega2masas}
\end{eqnarray}
where $T_m$ is the torque applied by the blades, $\omega_t$, is the speed of the blade-side shaft, $D_t$ and $H_t$ are parameters of the blade-side mass, $T_e$ is the electrical torque applied by the \gls{DFIG} on the shaft, $\omega_r$ is the \gls{DFIG}-side speed, $H_g$ and $D_g$ are the parameters of the \gls{DFIG}-side mass, $D_{tg}$ is the friction coefficient between the two masses, and $T_{tg}$ is the spring torque between the two masses.  All the parameters of this model and its conversion in $pu$, as used in (\ref{eq.omega1masas})-(\ref{eq.omega2masas}), are in Table~\ref{tab:2-mass-param} of the Appendix.

%---------------------------------------------------------
%\paragraph*{Control architecture}
%--------------------------------------------------------- 
\vspace{-2mm}
\subsection{Control architecture}
\begin{figure*}[t] 
  \centering
  \includegraphics[width=\textwidth,page=1]{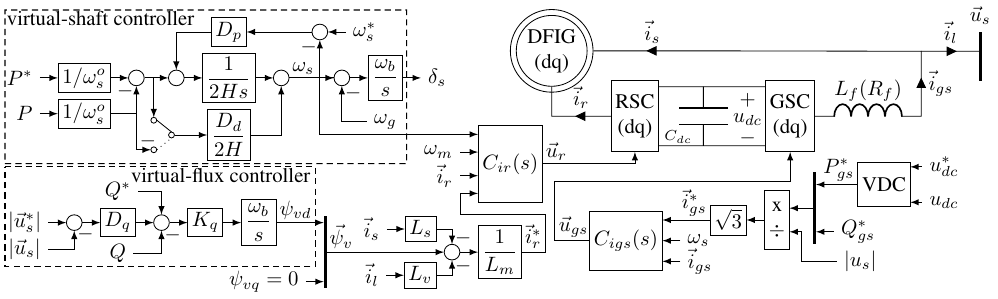}
  \caption{Hierarchical control structure of a VSM-controlled DFIG generating unit.}
  \label{fig:controlStructure-general}
\end{figure*}

\begin{figure}[t] 
  \centering
  \begin{subfigure}{0.48\linewidth}
    \centering
    \includegraphics[page=2,width=\linewidth]{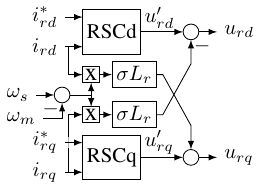}
    \caption{RSC current controller.}
    \label{fig:controlDiagram-RSC}
  \end{subfigure}
  \hfill
  \begin{subfigure}{0.48\linewidth}
    \centering
    \includegraphics[page=3,width=\linewidth]{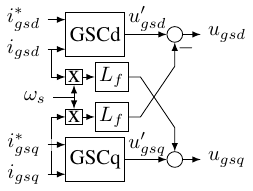}
    \caption{GSC current controller.}
    \label{fig:controlDiagram-GSC}
  \end{subfigure}
  \caption{Current-control loops of the DFIG.}
  \label{fig:controlStructure-RSC-GSC}
\end{figure}

A comprehensive five-layer control hierarchy (see Fig.~\ref{fig:controlStructure-general}) was implemented on each \gls{DFIG} unit. It includes:
\begin{itemize}
  \item[(a)] A \emph{Virtual-Shaft Controller} (\gls{VSM}), which emulates synchronous machine inertia and damping characteristics for primary frequency support.
  \item[(b)] A \emph{Virtual-Flux Controller}, responsible for regulating the magnitude of the \gls{DFIG} stator flux.
  \item[(c)] Inner \emph{Rotor Current Controllers} $C_{ir}(s)$ for the \gls{RSC}, managing the $d$- and $q$-axis rotor currents. Its internal structure is in Fig.~\ref{fig:controlDiagram-RSC}.
  \item[(d)] A \emph{DC-Link Voltage Controller}, $C_{V\!dc}(s)$ for  the DC bus between the \gls{RSC} and \gls{GSC}.
  \item[(e)] Inner \emph{Grid-Side Current Controllers}, $C_{igs}(s)$, for the \gls{GSC}, regulating the $d$- and $q$-axis currents injected into the grid. Details are in Fig.~\ref{fig:controlDiagram-GSC}.
\end{itemize}
For the scope of this study, reactive-power injection at the \gls{PCC} by the \gls{GSC} was disabled ($Q_{gs}^{\star}=0$).

%---------------------------------------------------------
%\paragraph*{Two-degree-of-freedom PI regulators}
%---------------------------------------------------------

\noindent \textbullet~{\em Two-degree-of-freedom PI regulator (2-DOF)}: 
The inner-loop controllers— specifically, the $dq$-axis current controllers for both the \gls{RSC} ($\text{RSC}_{dq}$) and \gls{GSC} ($\text{GSC}_{dq}$), as well as the DC-link voltage controller (VDC)— take the general form of:
\begin{equation}
  C_{\mathrm{PI}}(s)=K_{p}\bigl(b\,r-y\bigr)+\frac{K_{i}}{s}\bigl(r-y\bigr),
  \label{eq:2dofPI}
\end{equation}
where $r$ is the set point for the controlled variable, $y$ is its measured feedback value, $b$ is the set-point-weighting factor, and $K_{p}$ and $K_{i}$ are the proportional and integral gains, respectively. 
%PI structures are illustrated in Fig.~\ref{fig:pi-controller-config}.

%---------------------------------------------------------
%\paragraph*{Virtual synchronous machine implementation}
%---------------------------------------------------------
\noindent \textbullet~{\em \gls{VSM} implementation}:
The \gls{VSM} control logic implemented herein follows the "Alternative~I" configuration detailed in~\cite{GonzalezCajigas2023}, but incorporating two key refinements:% for enhanced applicability and performance:
\begin{itemize}
  \item[(a)] A full \gls{pu} formulation for all \gls{VSM} parameters and signals.
  \item[(b)] An additional damping term, $D_{d}$,  applied directly to the virtual shaft torque, as proposed in~\cite{Roldan2019}, to improve the damping of electromechanical oscillations.
\end{itemize}
 The scaling factor $\omega_b$ (base angular frequency in rad/s), in certain integrators within the \gls{VSM} control diagram (Fig.~\ref{fig:controlStructure-general}), is a requisite of the \gls{pu} implementation. 

%-------- FIGURE BLOCKS --------

%=========================================================
% End of Section II
%=========================================================
%=========================================================
\section{Initial Control Design and Baseline Analysis}
\label{sec:initialControlAnalysis}
%=========================================================
\subsection{Design Assumptions for Sequential Tuning}
\label{subsec:design_assumptions_sequential}
The initial tuning of every control loop within the \gls{DFIG}   \gls{VSM} hierarchy was performed using a classical sequential design procedure, i.e., treating each loop in isolation ignoring any interaction. This approach is common in industry~(see \cite{Abad2011,GonzalezCajigas2023,Roldan2019}, for example). More precisely: 
 \begin{enumerate}[label=(\alph*), leftmargin=15pt]
  \item {\em No loop interaction}: Inner control loops (e.g., current controllers) are presumed to be significantly faster than outer loops (e.g., power or voltage controllers) and are thus considered to have an infinitely fast response when tuning the outer loops. Conversely, outer loops are treated as providing constant set points during the tuning of inner loops. Furthermore, each of the four \gls{DFIG} units is tuned independently, without considering inter-machine coupling. 
  \item {\em Operating Point}: The controllers are tuned for the case when the grid absorbs \(0.7\,\text{pu}\) of the wind farm rated active power, and with \(\mathit{SCR}=1\), at the PCC.
  \item {\em Reference Design Rules}: Furthermore,
        \begin{enumerate}[label=(\roman*), leftmargin=15pt]
          \item The virtual-shaft (VSMP) and virtual-flux (VSMQ) controllers followed the design in~\cite{Roldan2019, GonzalezCajigas2023}.
          \item The virtual impedance, part of the VSM strategy, was selected according to the methodology presented in~\cite{GonzalezCajigas2023}.
          \item Every \gls{PI} in  the rotor- and grid-side current controllers ($\text{RSC}_{dq}$, $\text{GSC}_{dq}$), and the DC-link voltage controller (VDC) was tuned using the pole placement technique. In~\eqref{eq:2dofPI}  \(b=1\)  was used. Neglecting any controller interaction, 
          %Under the conventional simplifying assumptions of decoupled plant dynamics for each loop, 
          this tuning approach aimed to achieve a canonical second-order closed-loop transfer function for each \gls{PI}-controlled subsystem, as discussed in  texts like~\cite{Abad2011}.% for \gls{DFIG} control.
        \end{enumerate}
\end{enumerate}

\subsection{Initial controller specifications and ideal open-loop metrics}
\label{subsec:initial_specs_metrics}
%---------------------------------------------------------
Table~\ref{tab:controlSpecs-ini} includes the performance specifications for each control loop, as defined during the ideal sequential design phase, where $t_{s}^{2\%}$ is the $2\%$ settling time, $\zeta$ is the desired damping ratio, and $\omega_n$ is the equivalent closed-loop natural frequency (all these in the ideal second-order closed-loop system). The open-loop stability margins are $\varphi_m$ (phase margin) and $\omega_o$ (gain-crossover frequency).  

\begin{table}[ht]
    \centering
    \caption{Initial  Specifications.}
    \label{tab:controlSpecs-ini}
    \begin{tabular}{l|cc|c|cc}
    \hline\hline
    & \multicolumn{2}{c|}{Specs} & CL & \multicolumn{2}{c}{OL} \\
    Control & $t_s^{2\%}$ (s) & $\zeta$ & $\omega_n$ (rad/s) & $\varphi_m$ ($^{\circ}$) & $\omega_o$ (rad/s) \\ \hline
    VSMP        & 1     & 0.707 & 5.65  & 67.2 & 8.07  \\
    VSMQ        & 2     & ---   & ---   & 90.0 & 2     \\
    RSC\(_{dq}\) & 4 ms  & 0.707 & 1414  & 65.7 & 2184  \\
    VDC         & 80 ms & 0.707 & 70.4  & 65.5 & 109.9 \\
    GSC\(_{dq}\) & 4 ms  & 0.707 & 1414  & 65.7 & 2184  \\ \hline\hline
    \end{tabular}
\end{table}

%=========================================================
% end of Section III
%=========================================================
%========================================================
%  SECTION IV  –  Interaction analysis
%========================================================
\vspace{-2mm}
\section{Interactions among controllers}
\label{sec:interactions}

The primary objective of this section is to rigorously test the validity of the simplifying assumptions in Section~\ref{sec:initialControlAnalysis}. To this end, we have developed and applied a frequency-domain framework designed to: (i) isolate each individual controller within the complete, coupled multi-machine system; (ii) derive its true local open-loop transfer function, accounting for all interactions; and (iii) quantify the strength of the dynamic coupling between different control loops. 

%---------------------------------------------------------
\vspace{-2mm}
\subsection{General multi-loop control formulation}
\label{subsec:general_multiloop_formulation}
\begin{figure}[ht] % Figura original, sin cambios
    \centering    \includegraphics{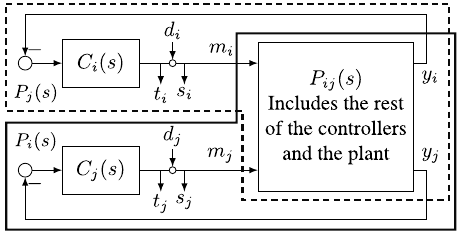}
    \caption{Multiple-controller diagram for interaction analysis.}
    \label{fig:multiple-control-diagram}
    \vspace{-3mm}
\end{figure}

%---------------------------------------------------------
Figure~\ref{fig:multiple-control-diagram} illustrates a generic $m$-controller architecture with notation and framework used in established multi-loop control system design texts~\cite{Skogestad2005}.  Controllers $C_i$ and $C_j$  manipulate variables $m_i$ and $m_j$ respectively (inputs to plant block $P_{ij}$). $P_{ij}$ encapsulates the physical subsystem and \emph{all} other active controllers in the system, thus embedding the inherent couplings. Auxiliary perturbation signals $d_i$ and $d_j$ are introduced in the controller outputs, and their corresponding effects are measured at sensor outputs $(s_i,t_i)$ and $(s_j,t_j)$. This setup allows for the computation of standard sensitivity $S_i(s)=s_i(s)/d_i(s)$ and complementary-sensitivity $T_i(s)=-t_i(s)/d_i(s)$ functions for each loop $i$, while the other loops remain closed, without altering the original closed-loop control structure:
\begin{equation}
  S_i(s)  = \frac{1}{1 + C_i(s)P_i(s)}\;\;\&\;\; 
  T_i(s)    = \frac{C_i(s)P_i(s)}{1 + C_i(s)P_i(s)},    \label{eq:S_i}
\end{equation}

$P_i(s)$ represents the effective plant \emph{as seen} by controller $C_i$ when its input is $m_i$ and its output is $y_i$. The local open-loop transfer function for loop $i$, $G_i(s)=C_i(s) P_i(s)$ is $T_i(s)/S_i(s)$. % obtained as: 
\begin{comment}
\begin{equation}
  G_i(s) = C_i(s)P_i(s) = \frac{T_i(s)}{S_i(s)}.
  \label{eq:G_i}
\end{equation}
\end{comment}
A control loop $i$ is considered \emph{robustly decoupled} if $G_i(s)$ has the specified gain and phase margin targets, even while all other controllers in the system remain active and interacting.

%---------------------------------------------------------
\vspace{-4mm}
\subsection{Obtaining 1-DOF Controllers for Loop Shaping}
\label{subsec:oneDOF}
%---------------------------------------------------------
Standard frequency-domain loop-shaping techniques are most directly applied to {\em 1-DOF} controllers. Therefore, the {\em 2-DOF} \gls{PI} structures used in the \gls{DFIG} control hierarchy must be reconfigured for this analysis.

\noindent \textbullet~{\em PI current and DC-link loops}
For the 2-DOF PI controllers  used in the current control loops ($\text{RSC}_{dq}$, $\text{GSC}_{dq}$) and the DC-link voltage control (VDC), 
%the set-point-weighting constant 
$b$ in~\eqref{eq:2dofPI} was fixed at $b=1$. %This effectively converts the 2-DOF structure into the equivalent 1-DOF form shown in Fig.~\ref{fig:pid-1dof}, which is suitable for standard loop analysis.

\begin{figure}[ht] 
\vspace{-3mm}
    \centering
    \includegraphics[width=0.5\textwidth]{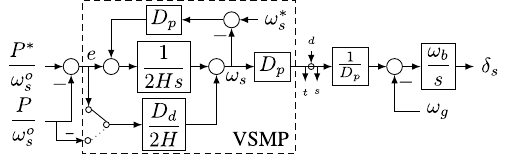}
    \caption{Virtual-shaft controller (VSMP) with  1-DOF structure. %The droop $D_p$ is factored out, leaving $C_{\text{VSMP}}(s)$ with unit DC gain.
    }
    \label{fig:VSMP-manipulation}
\end{figure}

\noindent \textbullet~{\em Virtual-shaft controller (VSMP)}
The \gls{VSM} power-frequency loop (VSMP) incorporates a steady-state droop characteristic ($D_p^{-1}$), which is typically mandated by grid codes and thus was considered a fixed parameter. To facilitate loop shaping, this droop gain was factored out of the main control loop, as illustrated in Fig.~\ref{fig:VSMP-manipulation}, and the term $D_d/(2H)$ was connected to the error between $P^*/\omega_s^0$ and $P/\omega_s^0s$ via a switch, which can later be changed to use only $-P/\omega_s^0$. The remaining portion within the dashed box constitutes a 1-DOF controller $C_{\text{VSMP}}(s)$, while the overall steady-state gain of the virtual-shaft power control path remains $D_p^{-1}$,
\begin{equation}
    C_{\text{VSMP}}(s)=
    \frac{1 + sD_d}{1 + s\!\left(2H/D_p\right)}.
  \label{eq:C_vsmp}
\end{equation}

\noindent \textbullet~{\em Virtual-flux $Q$ controller (VSMQ)}
Similar to the VSMP loop, the droop $D_q$ in the virtual-flux reactive power controller (VSMQ) is a fixed parameter. Consequently, only the proportional gain $K_q$ is a design variable for loop shaping. The output of the integrator, $(1/\omega_b)(d\psi_{vd}/dt)$, was taken as the manipulated variable $m$ and the integrator block $\omega_b/s$ was considered part of the plant, revealing an explicit 1-DOF proportional controller $K_q$.

%---------------------------------------------------------
\vspace{-2.5mm}
\subsection{Computing local plants and actual open-loop margins}
\label{subsec:computing_local_plants}
%---------------------------------------------------------
Let us now determine, for every control loop within the DFIGs: (i) the actual plant transfer function $P_i(s)$ that each controller $C_i(s)$ ``sees'' when all system interactions are present, and (ii) the true open-loop stability margins (phase margin and gain-crossover frequency) of each loop $G_i(s)$. This computation was carried out using \textsc{Matlab/Simulink} in four steps (Version R2024a or later is recommended for compatibility with the described linearization tools): 
%and involves the following four steps:

\begin{enumerate}[label=(\alph*), leftmargin=15pt]
  %-------------------------------------------------------
  \item \textit{Steady-State Operating Point}: The \verb|findop| command in SIMULINK was used to trim the complete non-linear model of the wind farm to the specific operating point defined in Section~\ref{sec:initialControlAnalysis} (i.e., \(P_{\text{grid}} = 0.7\)~pu, $\lvert v_{g}\rvert = 1$~pu, and \(\mathit{SCR}=1\)). Accurate steady-state conditions $\{x^\star,u^\star\}$ can be found if  a strict tolerance is used (e.g., \(\lVert\dot x^\star\rVert_2 < 10^{-8}\)).
  %, ensuring an accurate steady-state condition .
  %-------------------------------------------------------
  \item \textit{Full-Model Linearisation and Input/Output (I/O) Extraction}: All design perturbation points $d_i$ and their associated sensor output pairs $(s_i,t_i)$ (as depicted in Fig.~\ref{fig:multiple-control-diagram}) were declared as linear-analysis points  within the Simulink model. A \emph{single} execution of the \verb|linearize| command (e.g., \verb|linsys = linearize(model,io,op)| using the defined I/Os and the obtained operating point ``op'') produced the state-space realisation \(\bigl(A,B;C,D\bigr)\) of the \emph{entire} closed-loop multi-machine system. Subsequently, %for each controller $i$:
        \begin{enumerate}[label=(\roman*), leftmargin=14pt]
          \item The sensitivity path from $d_i$ to $s_i$ was extracted from the full linearised system \verb|linsys| (e.g., using index-based extraction like \verb|linsys(idx_s_i,idx_d_i)|) to obtain the sensitivity function $S_i(s)$ for each controller.
          \item Similarly, the complementary-sensitivity path from $d_i$ to $t_i$ was extracted to obtain $T_i(s)$.
          \item The local open-loop transfer function $G_i(s)$ was then computed using~\eqref{eq:S_i}.
          \item Since the 1-DOF controller $C_i(s)$ is known analytically (from Section~\ref{subsec:oneDOF}), the plant effectively ``seen'' by this controller was calculated as \(P_i(s)=G_i(s)/C_i(s)\).
        \end{enumerate}

  %-------------------------------------------------------
  \item \textit{Stability Margin Calculation}: The standard \verb|margin| command in MATLAB, when applied to each local open-loop transfer function $G_i(s)$ obtained above, yields the exact  \(\varphi_m\) and \(\omega_o\) for loop $i$.
\end{enumerate}

The actual $\varphi$ and $\omega_o$ derived from step (c) are presented as Stage~$\Beta$ in Table~\ref{tab:initialOLspecs-all}. For comparison, Stage~$\Alpha$ in the same table reiterates the original ideal design targets from Table~\ref{tab:controlSpecs-ini}, and Stage~\(\Gamma\) will later show the improved margins achieved after the coordinated redesign procedure detailed in Section~\ref{sec:designExample}.  For the time being, stage $\Beta$ of Table~\ref{tab:initialOLspecs-all} clearly reveals the inadequacy of the sequential tuning approach. Most notably, the $dq$-axis rotor-current loops (RSCd,q) see their phase margins collapse.
% AGC: Corregida la referencia a la sección de rediseño, asumiendo que sec:designExample es la V.

\begin{table}[ht] % Tabla original, verificar consistencia de ωo y referencias.
    \centering
    \caption{Open-loop characteristics (not from any industrial system): initial targets, actual under coupling, and after redesign. $\varphi_m$ in (\unit{\degree}), $\omega_o$ in rad/s.}
    \label{tab:initialOLspecs-all}
    \begin{tabular}{lccccccc}
    \hline\hline
                      & VSMP & VSMQ & RSCd & RSCq & VDC & GSCd & GSCq \\ 
    %%%%%%%%%%%%%%%%%%%%%%%%%%%%%%%%%%%%%%%%%%
    \hline 
    \multicolumn{8}{l}{$\Alpha$: Initial design targets (decoupled, from Table~\ref{tab:controlSpecs-ini})} \\ 
    $\varphi_m$ & \num{67.2} & \num{90.0} & \num{65.7} & \num{65.7} & \num{65.5} & \num{65.7} & \num{65.7} \\
    $\omega_o$  & \num{8.07} & \num{2.0} & \num{2184} & \num{2184} & \num{109.9} & \num{2184} & \num{2184} \\
    %%%%%%%%%%%%%%%%%%%%%%%%%%%%%%%%%%%%%%%%%%
    \hline
    \multicolumn{8}{l}{$\Beta$: Actual margins of initial design (coupled, from Section~\ref{subsec:computing_local_plants})} \\ % Corrected reference
    $\varphi_m$ & \num{86.6} & \num{105.0} & \num{15.7} & \num{14.7} & \num{64.5} & \num{47.0} & \num{42.3} \\
    $\omega_o$  & \num{12.4} & \num{2.7} & \num{332} & \num{339} & \num{122.7} & \num{1050} & \num{1045} \\
    %%%%%%%%%%%%%%%%%%%%%%%%%%%%%%%%%%%%%%%%%%
    \hline
    \multicolumn{8}{l}{$\Gamma$: Control margins after coordinated redesign (Section~\ref{sec:designExample})} \\ % Corrected reference
    $\varphi_m$ & \num{61.3} & \num{104.3} & \num{65.4} & \num{64.4} & \num{65.5} & \num{64.8} & \num{54.5} \\
    $\omega_o$  & \num{8.7} & \num{2.1} & \num{2318} & \num{2098} & \num{109.9} & \num{2170} & \num{2372} \\
    %%%%%%%%%%%%%%%%%%%%%%%%%%%%%%%%%%%%%%%%%%
    \hline \hline
    \end{tabular}
\end{table}

%---------------------------------------------------------
\vspace{-3mm}
\subsection{Interaction indices and derivation of redesign sequence}
\label{subsec:interaction_indices}
%---------------------------------------------------------

A quantitative metric is essential to capture how strongly each controller influences others and, conversely, how sensitive it is to disturbances propagating from other loops. The calculation of these interaction indices must proceed as follows:

\begin{enumerate}[label=(\alph*), leftmargin=15pt]
  \item A unit impulse perturbation is injected at the design input \(d_i\) of a specific loop \(i\), while all other perturbation inputs \(d_j\) (for $j \neq i$) are kept equal zero.
  \item The time-domain perturbation-rejection signals are recorded over a defined interval ($0$ to $t_{{\rm f}}$), typically $t_{{\rm f}}=1\,\mathrm{s}$: \(\sigma_i^{(i)}(t) \triangleq s_i(t)\) the response of loop $i$  to its own perturbation and \(\sigma_j^{(i)}(t) \triangleq s_j(t)\) (for \(j\neq i\)) the response of loop $j$ for the perturbation in loop $i$. 
\end{enumerate} 

The {\em directional correlation coefficient}, \(\rho_{ij}\), from the perturbed loop \(i\) to an observed loop \(j\) can be defined as:
\begin{equation}
  \rho_{ij} \;=\;
  \frac{\displaystyle
        \Bigl|\!\int_{0}^{t_\mathrm{f}}
        \sigma_j^{(i)}(t)\,\sigma_i^{(i)}(t)\,dt\Bigr|}
       {\displaystyle
        \sqrt{\int_{0}^{t_\mathrm{f}}
               \bigl[\sigma_j^{(i)}(t)\bigr]^2 dt}\,
        \sqrt{\int_{0}^{t_\mathrm{f}}
               \bigl[\sigma_i^{(i)}(t)\bigr]^2 dt}},
  \label{eq:rho_ki}
\end{equation}
where \(0 \le \rho_{ij}\le 1\). A large value of \(\rho_{ij}\) indicates that a disturbance primarily handled by loop \(i\) induces a significant time response in loop \(j\) strongly related with  the self-response of loop \(i\).

\begin{figure}[t] 
  \centering
  \includegraphics{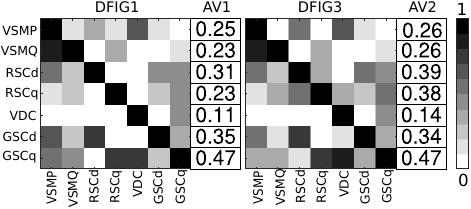}
  \caption{Absolute correlation matrix \(\boldsymbol\rho\) of perturbation–rejection (P–R) signals when each controller in DFIG1 (left) and DFIG3 (right) is perturbed. Darker cells mean stronger coupling. The columns ``AV1'' (``AV2'') show the average influence of each controller of DFIG1 (DFIG3) named in that row onto all the rest of the controllers in DFIG1 (DFIG3).}
  \label{fig:dfig1-controller-interaction}
\end{figure}

Repeating this impulse-response experiment for every possible excitation index \(i\) allowed the construction of the square interaction matrix \(\boldsymbol\rho = [\rho_{ij}]\). This matrix is shown as a grey-scale heat map in Figure~\ref{fig:dfig1-controller-interaction}, where darker cells are associated with stronger coupling. Each row in Figure~\ref{fig:dfig1-controller-interaction} shows how each controller in DFIG1 or DFIG3  interacts with the other controllers inside the same DFIG, using  these correlations. The average of that interaction is summarised in column AV1 (for DFIG1) and AV2 (for DFIG3). The fundamental interaction mechanisms remain similar.

%---------------------------------------------------------
%---------------------------------------------------------

From the full coupling matrix \(\boldsymbol\rho\), two scalar metrics can be extracted for each control loop to summarize its global impact on the system and its susceptibility to disturbances from other loops. Here, $n=7$ is the number of primary control loops per DFIG, namely: the virtual-shaft power (VSMP) and virtual-flux (VSMQ) controllers, the d- and q-axis rotor current controllers (RSCd, RSCq), the DC-link voltage controller (VDC), and the d- and q-axis grid-side current controllers (GSCd, GSCq). Two indices will be of interest here: 
(a) \emph{Influence Index (IIdx)} or the off-diagonal element average in row $i$ of $\boldsymbol\rho$ %(i.e., $\rho_{ij}$ for $j \neq i$) 
that quantifies how strongly control loop $i$ (when perturbed) affects all \emph{other} control loops $j$ and (b) \emph{Sensitivity Index (SIdx)} or the off-diagonal element average in column $i$ of $\boldsymbol\rho$ %(i.e., $\rho_{ji}$ for $j \neq i$). It 
that quantifies how strongly control loop $i$ is affected \emph{by} disturbances originated in all other control loops $j$: 

    \begin{equation}
%  \bar\rho_i^{\mathrm{out-influence}}
  IIdx = \frac{1}{n-1}
    \sum_{\substack{j=1,\,j\neq i}}^{n}
      \rho_{ij}\;\;\&\; \; SIdx  = \frac{1}{n-1}
    \sum_{\substack{j=1, \, j\neq i}}^{n}
      \rho_{ji}
\label{eq:influence_index_exc_diag_renamed}
\end{equation}
\begin{comment}
    (Note:  The columns "av1", "av2", "av3" in Fig.~\ref{fig:dfig1-controller-interaction} represent this influence of DFIG1 controllers on DFIG1 (other loops), DFIG2, and DFIG3, respectively.)

\emph{Sensitivity Index (SIdx)}: 
This is the average of the off-diagonal elements in column $i$ of $\boldsymbol\rho$ (i.e., $\rho_{ji}$ for $j \neq i$). It quantifies how strongly control loop $i$ is affected \emph{by} disturbances originating in all other control loops $j$:
\begin{equation}
%  \bar\rho_i^{\mathrm{in-sensitivity}}
SIdx  = \frac{1}{n-1}
    \sum_{\substack{j=1, \, j\neq i}}^{n}
      \rho_{ji}.
\label{eq:sensitivity_index_exc_diag_renamed}
\end{equation}
\end{comment}

Excluding the diagonal element $\rho_{ii}$ in both summations ensures that a controller self-correlation (response to its own perturbation) is not counted when assessing its interaction with other loops.

%---------------------------------------------------------
%\paragraph*{Prioritized Redesign Sequence}
%---------------------------------------------------------
 
\subsection{Prioritized redesign sequence} \label{sec.prioridades}
The redesign of controllers should prioritize those loops that exert the strongest overall influence on the rest of the system, as mitigating their disruptive impact first can simplify the subsequent tuning of other, less influential or more sensitive, loops. By sorting the average {\em IIdx}   values for DFIG~1 and DFIG~3 controllers (e.g., average of “AV1” and "AV2" columns in Fig.~\ref{fig:dfig1-controller-interaction}) in descending order, we derive the following optimal redesign sequence that will be followed in Section~\ref{sec:designExample}:
\begin{equation}
%\begin{gathered}
 {\small  \mathrm{GSC}_q;\,
  \mathrm{GSC}_d;\,
  \mathrm{RSC}_d;\,
  \mathrm{RSC}_q;\,
  \mathrm{VSMP};\,
  \mathrm{VSMQ};\,
  \mathrm{VDC}}
%\end{gathered}
\label{eq:design_seq}
\end{equation}
%---------------------------------------------------------
\vspace{-5mm}
\subsection{Time‐Domain Confirmation of Interactions}
\label{subsec:time_domain_confirmation}
\begin{figure}[t]
  \centering
    \includegraphics[height=6cm,width=0.4\textwidth]{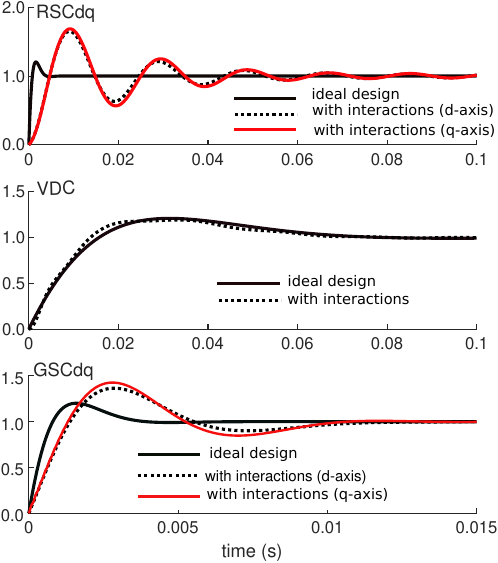}
  \caption{Unit‐step responses.  Ideal responses (Stage \(\Alpha\) vs actual responses.  (Stage \(\Beta\)).} 
  \label{fig:TRD-comparison-interactions}
  \vspace{-7mm}
\end{figure}

%---------------------------------------------------------
To corroborate the frequency‐domain findings regarding controller interactions and stability-margin degradation, step‐response tests were conducted on each controller  reference input using the full linearised model. Figure~\ref{fig:TRD-comparison-interactions} compares two scenarios:
\begin{enumerate}[label=(\alph*), leftmargin=15pt]
  \item Stage \(\Alpha\) (dark-grey traces): The ideal step responses predicted by the ideal decoupled second-order reference models
  %used during the initial sequential tuning 
  (as per Table~\ref{tab:controlSpecs-ini}).
  \item Stage \(\Beta\) (black-dashed and light-grey traces): The actual step responses obtained from the full linearised model of the wind farm.
\end{enumerate}

Fig.~\ref{fig:TRD-comparison-interactions} and Table~\ref{tab:initialOLspecs-all} (comparing Stage \(\Alpha\) and \(\Beta\)) show that:
% (Las observaciones (i) a (vi) se mantienen como en el original, ya que son sólidas y directamente relacionadas con los resultados)
\begin{enumerate}[label=(\roman*), leftmargin=15pt]
    \item \textit{Conservative VSMP design}: Under full coupling (stage \(\Beta\)), the VSMP loop actually shows lower overshoot and a faster settling time than originally specified (Table~\ref{tab:initialOLspecs-all}, \(\Beta\) vs.\ \(\Alpha\)).  Its phase margin increases from \(67.2^\circ\) to \(86.6^\circ\), and the bandwidth grows (from 8.07 to 12.4\,rad/s), indicating that the initial tuning was conservative and that this loop was more robust than required. 
    %when inner‐loop dynamics are included.
    \item \textit{Directional coupling from GSC\(_{dq}\) to RSC\(_{dq}\)}: The heat map in Fig.~\ref{fig:dfig1-controller-interaction} shows markedly darker cells for the paths GSC\(_{dq}\)\(\to\)RSC\(_{d,q}\) than for the reverse direction, indicating that the grid‐side converter current control strongly affects the rotor‐side loops.  Under full interaction (stage \(\Beta\)), this coupling produces a pronounced 50 Hz oscillation in the RSC\(_{dq}\) step response (Fig.~\ref{fig:TRD-comparison-interactions}), and corresponds quantitatively to the collapse of RSC\(_{dq}\) phase margin (Table~\ref{tab:initialOLspecs-all}, \(\Beta\) vs.\ \(\Alpha\)). % quitamos eso por ahora %%% This asymmetric interaction is the root cause of the severe damping degradation observed in the rotor‐current loops. 
    \item \textit{Moderate damping loss in \gls{GSC} loops}:  The \gls{GSC} current loops (GSC\(_{d,q}\)) exhibit a phase‐margin reduction from approximately 65° to around 42-47° (Table~\ref{tab:initialOLspecs-all}, \(\Beta\) vs.\ \(\Alpha\)), consistent with the weaker coupling in the reverse direction (RSC\(\to\)GSC) visible in Fig.~\ref{fig:dfig1-controller-interaction}. 
    This further confirms the asymmetric interaction between rotor‐ and grid‐side current controls.
    \item \textit{Robustness of VSMQ and VDC loops}: The virtual‐flux controller (VSMQ) actually gains phase‐margin under full interaction, increasing from 90° to 105° (Table~\ref{tab:initialOLspecs-all}, \(\Beta\) vs.\ \(\Alpha\)). %, and its step response (Fig.~\ref{fig:TRD-comparison-interactions}, second panel) shows only a slight change in rise time. 
    The DC‐link controller (VDC) similarly retains a high phase margin, dropping marginally from 65.5° to 64.5°, and its time‐domain response (Figure~\ref{fig:TRD-comparison-interactions}) remains well within the original design envelope. %These results, 
    %combined with the relatively light correlation coefficients in Fig.~\ref{fig:dfig1-controller-interaction} (influence indices generally below 0.34 for VSMQ and 0.23 for VDC when perturbed), 
    %confirm that 
    Both VSMQ and VDC loops are inherently robust, and largely decoupled from the more critical interactions affecting RSC\(_{dq}\) and GSC\(_{dq}\).
  \item \textit{Methodology validation}: The strong alignment between: (a) the analytically predicted stability margin erosion (Table~\ref{tab:initialOLspecs-all}, Stage \(\Beta\)), (b) the quantified coupling strengths from the correlation matrix (Fig.~\ref{fig:dfig1-controller-interaction}), and (c) the observed time‐domain performance degradation (oscillations and deviations from ideal responses in Fig.~\ref{fig:TRD-comparison-interactions})  demonstrates the consistency and predictive power of the proposed multi‐loop, interaction‐based analysis framework. 
  %No single analysis technique in isolation (neither purely frequency‐domain calculations based on decoupled models nor purely time‐domain simulations without insight into loop stability) could have revealed this complete picture of system interactions.
  \item \textit{Need  for coordinated redesign}: %The substantial degradation of stability margins and time-domain performance, particularly for the RSC and GSC loops, when moving from the idealized Stage \(\Alpha\) to the fully coupled Stage \(\Beta\), underscores the  shortcomings of sequential and decoupled tuning for this \gls{VSM}-\gls{DFIG} system. %The systematic approach to identifying influential loops via the derived 
  The redesign sequence in~\eqref{eq:design_seq} provides a clear path towards performance recovery, as shown afterwards. Unlike a formal $\mu$-analysis, the current framework identifies and quantifies dominant interaction pathways based on the nominal system parameters.
  %, which will be demonstrated by the improved margins in Stage \(\Gamma\) of Table~\ref{tab:initialOLspecs-all} following the coordinated redesign.
\end{enumerate}

%========  END OF SECTION IV =========

\vspace{-4mm}
\section{Coordinated Redesign: Methodology Example}
\label{sec:designExample} % Cambiado de sec:control-redesign si este es el nuevo label

This section illustrates the proposed coordinated frequency-domain redesign procedure. The methodology, which relies on the interaction analysis and the prioritised redesign sequence from Section~\ref{sec:interactions} %(specifically the prioritized redesign sequence in~\eqref{eq:design_seq}), 
has been applied iteratively to each controller. Here, we only detail the redesign of one critical loop as an example: the rotor-side \(q\)-axis current controller (RSC\(_q\)) of DFIG~1. The same systematic steps have be applied to all other controllers according to the derived priority order. The redesign of the RSC\(_q\) controller of DFIG~1 comprises the following steps:

\begin{enumerate}[label=(\alph*), leftmargin=15pt]
  \item \textit{Controller Selection and I/O Tagging for Analysis}:
        To analyse the RSC\(_q\) control block, specifically, the \(q\)-axis current loop,  (included in Fig.~\ref{fig:controlDiagram-RSC}) within the full system model, a linear-analysis point is inserted at its manipulated variable output (\(u'_{rq}\) in Fig.~\ref{fig:controlDiagram-RSC})  which serves as the perturbation injection point \(d\). The signals immediately before (controller output) and after (plant input, considering the summation node) these injection point are tagged as \(s\) and \(t\), respectively, for sensitivity function extraction.
        The nominal 1-DOF PI law for this controller ($b=1$ as in Section~\ref{subsec:oneDOF}) is given by:
        \begin{equation}
            C_{\mathrm{RSCq}}(s)
          = K_{p,\mathrm{RSCq}}
             + \frac{K_{i,\mathrm{RSCq}}}{s} 
    \label{eq:C_RSCq_nominal_V}
        \end{equation}
  \item \textit{Extraction of the Actual Plant and Current Open-Loop Response}:
        Using the complete linearized state-space model of the wind farm (derived in Section~\ref{subsec:computing_local_plants}), the sensitivity functions $S(s)$ and $T(s)$ for the RSC\(_q\) loop are computed,  and the current local open-loop transfer function $G(s)$ and the actual plant $P(s)$ seen by $C_{\mathrm{RSCq}}(s)$ can be determined:
        \begin{equation}
          G(s) = \frac{T(s)}{S(s)},
          \quad
          P(s) = \frac{G(s)}{C_{\mathrm{RSCq}}(s)}.
          \label{eq:RSCq_openloop_plant_V} % Renombrado label
        \end{equation}
  \item \textit{Redesign Targets}:
        The objective is to reshape $G(j\omega)$ to achieve the original design targets specified in Table~\ref{tab:controlSpecs-ini} (Stage $\Alpha$ of Table~\ref{tab:initialOLspecs-all}). For the RSC\(_q\) loop, these are a crossover frequency \(\omega_o = 2184\)~rad/s and a phase margin \(\varphi_m = 65.7^\circ\).
        To design the new PI controller, we first read the identified plant's frequency response $P(j\omega_o)$ (magnitude $A_p$ and phase $\phi_p$)   at the target crossover frequency \(\omega_o\), %(point~\emph{c} in Fig.~\ref{fig:example-RSCq-nichols}), 
        yielding: $A_p =  -10\,\mathrm{dB},\;\&\; \phi_p = -93.8^\circ$. 

\begin{comment}
\begin{equation}
          A_p =  -10\,\mathrm{dB},
          \;\&\; \phi_p = -93.8^\circ.
          \label{eq:plant_gain_phase_V} % Renombrado label
        \end{equation}
\end{comment}
  \item \textit{Computation of New PI Controller Parameters}:
        The new PI compensator, $C_{\mathrm{RSCq,new}}(s)$, must provide a gain contribution $\Delta A = -A_p = +10$\,dB and a phase $\Delta\phi = \varphi_m - (180^\circ + \phi_p) = 65.7^\circ - (180^\circ - 93.8^\circ) = -20.5^\circ$, which gives two equations ($\Delta A$ and $\Delta \phi$) with two unknowns ($K_{p,RSCq}$ and $K_{i,RSCq}$). 
  \item \textit{Verification of Redesigned Loop}:
        With the new PI parameters ($K_{p,\mathrm{RSCq,new}}$, $K_{i,\mathrm{RSCq,new}}$), the reshaped open-loop transfer function \(G_{\mathrm{rd}}(j\omega) = C_{\mathrm{RSCq,new}}(j\omega)P(j\omega)\) has phase margin and crossover frequency equal to the specifications. 
\end{enumerate}

\begin{comment}
\begin{figure}[t]
  \centering
  \psfragfig{figures/designExample/design-example-RSCq}{
  }
  \caption{Nichols plot for the RSC\(_q\) loop redesign: (thick-grey line) identified plant \(P(j\omega)\); (thin-grey line) original open‐loop \(G(j\omega)\) with initial tuning (point \emph{a} shows its stability margins); (dark-grey line) redesigned open‐loop \(G_{\mathrm{rd}}(j\omega)\) (point \emph{b} shows its new margins). Point \emph{c} represents the plant response at the target crossover frequency \(\omega_o\).}
  \label{fig:example-RSCq-nichols}
\end{figure}
\end{comment}
\begin{comment}
\begin{table}[ht]
  \centering
  \caption{Results for the RSC\(_q\) loop redesign. \color{blue}QUITAR\color{black}}
  \label{tab:design-example-characteristics}
  \begin{tabular}{l|c c c c}
    \hline\hline
    Point & \(A\) (dB) & \(\phi\) (°) & \(\varphi_m\) (°) & \(\omega_o\) (rad/s) \\ 
    \hline
    a (Initial OL) &  0.0  & -165.3 & 14.7 & 339 \\ % Margins of G(s)
    b (Redesigned OL) &  0.0  & -114.3 & 65.7 & 2184 \\ % Margins of G_rd(s)
    c (Plant at $\omega_o$) & -10.0 & -93.8  &  ---   & 2184 \\ % P(jωo)
    \hline\hline
  \end{tabular}
\end{table}
\end{comment}

The same multi-step procedure was systematically applied to all seven controllers of each DFIG in the wind farm, following the priority order established by the influence analysis in Section~\ref{sec:interactions}~\eqref{eq:design_seq}. With only one iteration the results in   Table~\ref{tab:initialOLspecs-all} ($\Gamma$) are obtained, with stability margins close to the specifications. This iterative process may continue until all loops achieve their specified performance targets within the fully coupled system.

%%%%%%%%%%%%%%%%%%%%%%%%%%%%%%%%%%%%%%%%%%%%%%%%%%%%%%%%%%
% =========================================================
\vspace{-2mm}
\section{Results and validation}\label{sec:results}
% ---------------------------------------------------------
\subsection{Frequency-domain performance of redesigned controllers}
\label{subsec:freq-domain-design-results} % Label renombrado para unicidad
% ---------------------------------------------------------

The open-loop stability margins (phase margin $\varphi_m$ and crossover frequency $\omega_o$) achieved after the \textit{first full iteration} of this coordinated redesign process are reported in the $\Gamma$~row of Table~\ref{tab:initialOLspecs-all}. A direct comparison of these achieved margins (Stage~$\Gamma$) with the original ideal targets (Stage~$\Alpha$) and the severely degraded margins of the initial sequential (uncoupled) design (Stage~$\Beta$) confirms the effectiveness of the proposed methodology. Further redesign iterations should progressively drive the $(\varphi_m,\omega_o)$ pairs even closer to the set points specified in Stage~$\Alpha$. In this example, the second iteration yielded errors below 3\% in all specification targets.

\begin{comment}
\begin{itemize}
  \item All seven control loops now meet their target phase margins~$\varphi_m$ and crossover frequencies~$\omega_o$ (as specified in Stage~$\Alpha$) within narrow deviation bands% of~$\pm\,17.2\%$ for $\varphi_m$ and $\pm\,9.2\%$ for $\omega_o$, respectively, 
   after just one coordinated redesign iteration. This is a significant improvement over the initial sequentially tuned design.% (Stage~$\Beta$), where the deviations from target for the actual coupled system were as large as~$\pm\,77.6\%$ for $\varphi_m$ and $\pm\,84.8\%$ for $\omega_o$.
  \item Further redesign iterations should progressively drive the $(\varphi_m,\omega_o)$ pairs even closer to the precise set points specified in Stage~$\Alpha$.
\end{itemize}
\end{comment}

While this study focuses on achieving nominal performance targets, a formal robust stability analysis (e.g., using $\mu$-analysis or Monte Carlo simulations with parameter uncertainties) would be a valuable extension for industrial applications and is considered for future work (Section~\ref{sec:conclusions}).

\vspace{-4mm}
\subsection{Simulation of small- and large-disturbance response}
\label{subsec:small-signal-verification} % Label renombrado
% ---------------------------------------------------------

\begin{figure}[t]
    \centering
    \includegraphics[height=4.5cm,width=0.4\textwidth]{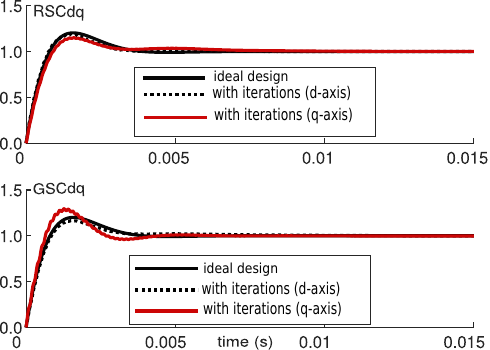}
    \caption{After redesign. Unit-step responses,  ideal 2nd-order templates, stage \(\Alpha\), vs obtained responses, stage \(\Gamma\).}
    \label{fig:FRD-comparison-interactions}
    \vspace{-5mm}
\end{figure}

Figure~\ref{fig:FRD-comparison-interactions} illustrates the unit-step responses for each primary control loop, comparing:
\begin{enumerate}[label=(\alph*)]
  \item the ideal second-order transient response %templates used during the controller specification phase 
  (Table~\ref{tab:controlSpecs-ini}), and
  \item the actual responses %obtained from the linearised multi-loop model 
  after implementing the controllers from the first coordinated redesign iteration ($\Gamma$ of Table~\ref{tab:initialOLspecs-all}).
\end{enumerate}

The redesigned controllers (dotted black and light-grey solid lines) demonstrate excellent reference tracking performance, closely aligned with the theoretical second-order template curves (solid grey lines). Only the most problematic loops in Figure~\ref{fig:TRD-comparison-interactions} are shown here.  Small deviations are further attenuated and typically vanish after a second redesign cycle.

% ---------------------------------------------------------
%\subsection{Large-signal non-linear validation}
%%\label{subsec:large-signal-validation} % Label renombrado
% ---------------------------------------------------------

Time-domain simulations, with large-signal disturbances, were performed on the detailed non-linear Simulink model of the DFIG wind farm. For the remainder of the paper, FRD (Frequency-Response Design) refers to the controllers obtained through the coordinated methodology presented in this paper, while the TRD (Time-Response Design) refers to the controllers from the initial sequential tuning (Section~\ref{sec:initialControlAnalysis}). Each simulation scenario begins at the nominal operating point defined in Section~\ref{sec:initialControlAnalysis} (grid voltage~$1\,$pu, wind-farm active-power reference~$0.7\,$pu, $\mathrm{SCR}=1$).

\vspace{-3mm} 
\begin{figure}[H]
  \centering
\includegraphics[height=3.5cm,width=0.9\linewidth]{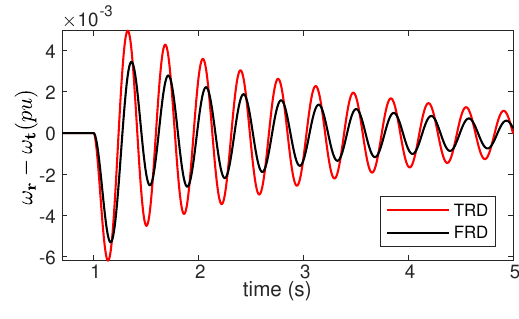}
  \caption{Rotor–turbine speed deviation \(\omega_r - \omega_t\) of DFIG~1 following a \(0.2\)\,pu step increase in \(P_{\mathrm{ref}}\) at \(t = 1\)\,s. Sequential Tuning (TRD); Coordinated Redesign (FRD).}
  \label{fig:nonLinSim_PREF_DT_STRAINANGLE}
  \vspace{2mm}
  \includegraphics[width=\linewidth]{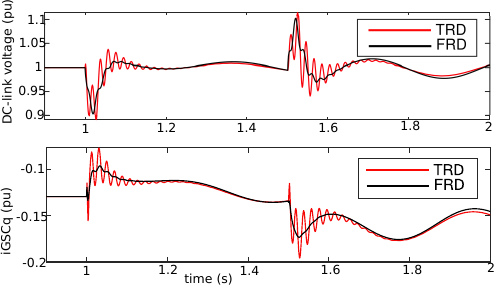}
  \caption{DC-link voltage and  \(q\)-axis grid-side converter current for DFIG~1 during a voltage-dip at the grid bus. Sequential Tuning (TRD) vs Redesign (FRD). Voltage dip ($0.2$\,pu at \(t = 1\,\mathrm{s}\) for $0.5$\,s)}
  \label{fig:nonLinSim_VDROP_iGSCdq}
\end{figure}

%----------------------------------------------------------
\subsubsection{Active-Power Reference Step ($\Delta P_{\mathrm{ref}} = 0.2$\,pu) in Gen\-erator~1, at $t=1\,s$}
\label{subsubsec:pref-step-results} % Label renombrado
%----------------------------------------------------------

Figure~\ref{fig:nonLinSim_PREF_DT_STRAINANGLE} illustrates the resulting rotor–turbine speed deviation (\(\Delta\omega = \omega_r - \omega_t\)). This deviation would produce a drivetrain torsional deformation. Notice that the FRD controllers reduce the initial overshoot and the final settling time. This should translate in reduced mechanical stress on the turbine drivetrain. 

%----------------------------------------------------------
\subsubsection{Voltage dip ($0.2$\,pu at \(t = 1\,\mathrm{s}\) for $0.5$\,s)}
\label{subsubsec:vdip-results} % Label renombrado
%----------------------------------------------------------

Figure~\ref{fig:nonLinSim_VDROP_iGSCdq} plots the DC-link voltage (\(u_{\mathrm{dc}}\)) of DFIG~1 (representative of other units), and the $q$-axis current of the \gls{GSC}. FRD controllers also reduce the DC-link and d-q oscillations. This should lead to a faster post-fault recovery.
% =======================================================
% END OF SECTION
% =======================================================
%========================================================
\vspace{-4mm} 
\section{Conclusions and Future Work}
\label{sec:conclusions}
%========================================================

This paper has introduced and validated a comprehensive \emph{frequency-domain, multi-loop} design methodology for the coordinated tuning of all critical control layers within a Virtual Synchronous Machine (VSM)-controlled Doubly-Fed Induction Generator (DFIG) wind farm. By  explicitly accounting for loop-to-loop and inter-machine dynamic interactions, the proposed approach  achieves the proposed designed goals and  high-performance operation under both small-signal and large-signal disturbances, even when subjected to the challenging conditions of very weak grids (e.g., SCR=1).

Results show that the systematic application of the perturbation-rejection framework (Section~\ref{sec:interactions})  identified and quantified loop interactions and revealed severe phase-margin erosion in critical loops like the RSC\(q\) when using conventional design techniques. However, a single iteration of the proposed coordinated redesign procedure restored all stability margins to almost their specified targets.  In addition, closed-loop step responses derived from the full small-signal model of the wind farm closely matched the ideal second-order performance templates defined during the design phase.  Finally, non-linear time-domain simulations with the full model of the wind farm confirmed the improved performance of the controllers tuned via the proposed approach with respect to the performance of those designed ignoring controller interactions. 

A significant advantage of the proposed methodology is its reliance on well-established frequency-domain metrics (gain/phase margins, Nichols diagrams) and standard linear system analysis tools. These are readily available in commercial simulation packages (e.g., MATLAB/Simulink).

From the theoretical viewpoint, further work is required to generalise the design procedure taking into account a rich set of possible operating points and to formulate an optimisation problem to systematically select individual loop crossover frequencies and phase margins. From the application viewpoint, testing the redesigned controllers in an experimental set-up will be pursued to assess the impact of real-world factors. 

\bibliographystyle{IEEEtran}
\vspace{-2mm}
%    \bibliography{RefPaper.bib,Retosbib.bib}
% Generated by IEEEtran.bst, version: 1.14 (2015/08/26)

%\pagebreak
\appendices 
\appendix
\setcounter{secnumdepth}{0}
%\section{System bases and model parameters}
\section{APPENDIX}
\label{apx:system-bases}
The system parameters and bases  are in Tables~\ref{tab:2-mass-param} and~\ref{tab:eachDFIGparameters}.

%\vspace{-5cm}
\begin{table}[!b]
    \centering
    \vspace{-5mm}
    \caption{Parameters of the 2-mass mechanical model}
    \label{tab:2-mass-param}
    \begin{tabular}{ccc}
       \hline \hline 
       parameter [units]                                   &  value       & conversion to pu model                   \\ \hline
       $J_t[kg \cdot m^2]$               &  \num{800}   & $H_t [\mathrm{s}] = J_t\omega_{m,b}^2/(2S_b)$   \\
       $D_t^\mathrm{r}[\mathrm{N\,m}\,\mathrm{s}\,\mathrm{rad}^{\mathrm{-1}}]$     &  \num{0.1}   & $D_t [\mathrm{pu}] = D_t^\mathrm{r}\omega_{m,b}^2/Sb$     \\
       $K_{tg}^\mathrm{r}[\mathrm{N\,m}\,\mathrm{rad}^{\mathrm{-1}}]$ &  \num{12500} & $K_{tg} [\mathrm{pu}] =K_{tg}^\mathrm{r}\omega_{m,b}/Sb$   \\
       $D_{tg}^\mathrm{r}[\mathrm{N\,m}\,\mathrm{s}\,\mathrm{rad}^{\mathrm{-1}}]$  &  \num{130}   & $D_{tg} [\mathrm{pu}] =D_{tg}^\mathrm{r}\omega_{m,b}^2/Sb$ \\
       $J_g[kg \cdot m^2]$               &  \num{90}    & $H_g [\mathrm{s}] = J_g\omega_{m,b}^2/(2Sb)$    \\
       $D_g^\mathrm{r}[\mathrm{N\,m}\,\mathrm{s}\,\mathrm{rad}^{\mathrm{-1}}]$     &  \num{0.1}   & $D_g [\mathrm{pu}] = D_g^\mathrm{r}\omega_{m,b}^2/S_b$    \\
       \hline \hline
\end{tabular}
\end{table}

\begin{table}[!b]
 \centering
\vspace{-2mm}
\caption{Parameters of each DFIG and base values 
($R_r$ and $L_r$ referred to stator). These data do not correspond to any specific commercial DFIG but to a generic 2~MW machine as in~\cite{Abad2011}}
\label{tab:eachDFIGparameters}
\begin{tabular}{ccl} \hline\hline
Parameter  & value (pu) & Description \\ \hline
$R_s/R_r$ & $6.6 \cdot 10^{-3}/7.4 \cdot 10^{-3}$ & Stat. Res./Rotor Res. \\
$L_s/L_r$ & 2.07/2.07 & Stat. Induct./Rot. Induct. \\ 
$L_M$ &  2.00 & Mutual Inductance \\
$u_{dc}$ & 2.00 & DC link voltage DFIG\\
$C_{dc}$&  0.485 & DC link capacitor \\\hline

\multicolumn{3}{c}{Base values for DFIGs and main grid} \\
\hline
$V_b$ & $690\,V$ & Phase-to-Phase nom. volt. \\
$I_b$ &  $1760\,A$ & Nom. current \\
$\psi_b$ & $V_b/\omega_b$ & Base flux linkage\\
$S_b$ & $\sqrt{3}V_b I_b = 2.1\,MVA$& Base Power \\
$Z_b$ & $V_b/I_b = 0.39\,\Omega$ & Base Impedance \\
$\omega_b$ & $2\cdot \pi \cdot 50 \;rad/s$ & Base angular freq. \\
$L_b$ & $Z_b/\omega_b$ & Base inductance \\
$\omega_{m,b}$ & $\omega_b/P$ in rad/s & Base angular rotor speed \\
$T_{e,b}$ & $S_b/\omega{m,b}$ in $N\cdot m$ & Base Torque \\
$J_b$ & $T_{e,b}/\omega_b$  & Base Mech. Inertia \\
$U_{dc,b}$ & $1200\,V$ & Base DC voltage \\
$I_{dc,b}$ & $S_b/U_{dc,b}= 1750\,A$ & Base DC current \\
$Z_{dc,b} $ & $U_{dc,b}/I_{dc,b} = 0.686\,\Omega$ & Base DC impedance \\
$C_{dc,b}$ & $1/(\omega_b Z_{dc,b})$ & Base DC capacitor
\\\hline
\end{tabular}
\end{table}
%%%%%%%%%%%%%%%%%%%%%%%%%%%%%%%%%%%%%%%%%%%%%%%%%%%%%%%%%%%%%%%%%%%%%%
%

%%%%%%%%%%%%%%%%%%%%%%%%%%%%%%%%%%%%%%%%%%%%%%%%%%%%%%%%%%%%%%%%%%%%%%%%%%%
\end{document}